\documentclass[singlecolumn,amssymb, nobibnotes, aps, prd,amsmath,mathtools, nofootinbib,pgfmath,pgffor,superscriptaddress]{revtex4-1}
\usepackage{graphicx}

\newcommand{\be}{\begin{equation}}
\newcommand{\ee}{\end{equation}}
\newcommand{\bw}{\begin{widetext}}
\newcommand{\ew}{\end{widetext}}
\newcommand{\ba}{\begin{eqnarray}}
\newcommand{\ea}{\end{eqnarray}}
\newcommand{\bi}{\begin{itemize}}
\newcommand{\ei}{\end{itemize}}

\newcommand{\bfi}{\begin{figure}
\epsfxsize=9cm
\epsffile}
\newcommand{\bfinew}{\begin{figure}
\begin{center}
\includegraphics}
\newcommand{\efi}{\end{figure}}
\newcommand{\efinew}{
\end{center}
\end{figure}}

\usepackage{hyperref}
\hypersetup{colorlinks,citecolor=blue}

\makeatletter
\def\qrr@split@result#1 #2\@qrr@split@result{\edef\erfInput{#1}\edef\erfResult{#2}}
\newcommand*{\gnuplotErf}[2][\jobname.eval]{%
    \immediate\write18{gnuplot -e "set print '#1'; print #2, erf(#2);"}%
    \everyeof{\noexpand}
    \edef\qrr@temp{\input{#1} }%
    \expandafter\qrr@split@result\qrr@temp\@qrr@split@result
}
\makeatother

\usepackage{multirow}
\newcommand{\beq}{\begin{equation}}
\newcommand{\eeq}{\end{equation}}
\newcommand{\bea}{\begin{eqnarray}}
\newcommand{\eea}{\end{eqnarray}}

\usepackage{hyperref}
\hypersetup{colorlinks,citecolor=blue}

\begin{document}

\title[Cosmological Zero Modes]{Cosmological Zero Modes}

\author{Niayesh Afshordi}
\email{nafshordi@pitp.ca}
\affiliation{Perimeter Institute for Theoretical Physics, 31 Caroline St. N., Waterloo, ON, N2L 2Y5, Canada}
\affiliation{Department of Physics and Astronomy, University of Waterloo, Waterloo, ON, N2L 3G1, Canada}

\author{Matthew C. Johnson}
\email{mjohnson@pitp.ca}
\affiliation{Perimeter Institute for Theoretical Physics, 31 Caroline St. N., Waterloo, ON, N2L 2Y5, Canada}
\affiliation{Department of Physics and Astronomy, York University, Toronto, Ontario, M3J 1P3, Canada}

\begin{abstract}
We introduce a new family of primordial cosmological perturbations that are not described by traditional power spectra. At the linear level, these perturbations live in the kernel of the spatial Laplacian operator, and thus we call them {\it cosmological zero modes}. We compute the cosmic microwave background (CMB) temperature and polarization anisotropy induced by these modes, and forecast their detection sensitivity using a  cosmic-variance limited experiment. In particular, we consider two configurations for the zero modes: The first configuration consists of stochastic metric perturbations described by white noise on a ``holographic screen'' located at our cosmological horizon. The amplitude of the power spectrum of this white noise can be constrained to be $\lesssim 9 \times 10^{-14}$. The second configuration is a primordial monopole beyond our cosmological horizon. We show that such a monopole, with ``charge'' $Q$, can be detected in the CMB sky up to a distance of $11.6 ~ Q^{1/4}\times$ horizon radius (or $160~ Q^{1/4}$ Gpc).  More generally, observational probes of cosmological zero modes can shed light on non-perturbative phenomena in the primordial universe, beyond our observable horizon. 
\end{abstract}

\maketitle

\section{Introduction}\label{sec:intro}

Cosmology has made tremendous progress over the past couple of decades. The simple six-parameter $\Lambda$CDM cosmological model is now able to fit nearly all observations, ranging from the cosmic microwave background (CMB) anisotropies to Lyman-$\alpha$ forest fluctuations in quasar spectra. This empirical success, however, has also highlighted gaping holes in our understanding of cosmos: the nature of dark matter, dark energy, and the cosmological big bang remain elusive.  

In the standard $\Lambda$CDM model, the early universe is extremely simple: It consists of a thermal plasma with nearly uniform expansion. The 3-geometry of a constant temperature surface in this era is described by a Euclidean geometry with small gaussian fluctuations, $\zeta$, in the conformal factor:
\beq
g_{ij}({\bf x})= e^{2\zeta({\bf x})} \delta_{ij}, \label{zeta}
\eeq
where, at early times, we have \cite{Ade:2015xua}:
\bea
&&\langle \zeta({\bf x}) \zeta({\bf y}) \rangle =  A_s \int \frac{d^3{\bf k}}{4 \pi k^3} e^{i {\bf k\cdot (x-y)}} \left( k \over 0.05 ~{\rm Mpc}^{-1} \right)^{n_s-1}, \nonumber \\
&&A_s = (2.195 \pm 0.079) \times 10^{-9}, n_s= 0.9645 \pm 0.0049. \label{power_spec}
\eea 

We should already notice that, since $n_s <1$ at 7$\sigma$ confidence level~\cite{Ade:2015xua}, the integral in Eq. \eqref{power_spec} has an infrared divergence. This means that there are no meaningful predictions for the amplitude of Fourier modes with $k^2 \rightarrow 0$ in the concordance $\Lambda$CDM cosmological model. 

Another way to illustrate the lack of constraining power on ultra large scales is to note that all cosmological observables depend on the probability distribution of $\zeta({\bf x})$ at early times (e.g., on the inflationary reheating surface). As all (primary) CMB observations are consistent with gaussian statistics \cite{Ade:2015ava}, the probability functional for $\zeta({\bf x})$ can be modelled by:
\beq
{\cal P}[\zeta] \propto \exp\left[-\int \zeta({\bf x}) F(\Delta) \zeta({\bf x}) d^3{\bf x} \right], \label{probability}
\eeq
where $\Delta$ is the spatial Laplacian operator, and
\beq
F(\Delta) = \frac{(-\Delta)^{2-n_s/2}}{2\pi^2 A_s (0.05 ~{\rm Mpc}^{-1} )^{1-n_s}},
\eeq
for a power-law power spectrum \eqref{power_spec}. 

From this point of view, as long $n_s  \leq 2$, the saddle point involving any moment of ${\cal P}[\zeta]$ (i.e. $\delta\ln {\cal P}/\delta\zeta =0$), satisfies:
\beq
F(\Delta)  \zeta =0~~~ \Rightarrow~~~ \Delta \zeta =0, \label{Laplace}
\eeq 
again implying that the modes in the kernel of the Laplacian operator are unconstrained. We call these {\it cosmological zero modes}.

We should note that the power-law nature of the power spectrum (or $F(\Delta)$) is only empirically verified for finite wavenumbers, and it certainly may not extend all the way to $k \rightarrow 0$. Past studies have focused on the observational implications of large-amplitude superhorizon perturbations (e.g., \cite{GarciaBellido:1995wz,Erickcek:2008jp}), often characterized as {\it Grichuk-Zel'dovich effect} \cite{1978SvA....22..125G}, or as the ultra-large scale structure of the universe \cite{1991ASSL..169..253S, Braden:2016tjn}.  Here, we contrast these long wavelength perturbations with the {\it zero modes}, which have formally infinite wavelength. As we see below, working in this limit leads to a distinct theoretical framework, allowing for significant simplifications and sharper observable predictions for the zero modes.

We first show that the existence of zero modes cannot be described by any analytic description of the power spectrum, and requires a non-perturbative model of cosmological perturbations in the early universe. To see this, we can write down the most generic solution to the Laplace equation \eqref{Laplace}, in spherical coordinates $(\chi,\theta,\phi)$, around an arbitrary origin:
\beq
\zeta(\chi,\theta,\phi) = \sum_{\ell,m} (A_{\ell m} \chi^\ell+ B_{\ell m} \chi^{-\ell-1}) Y_{\ell m}(\theta,\phi),
\eeq   
where $Y_{\ell m}$'s are the spherical harmonics. Therefore, we see that any zero mode should blow up at $\chi =0$ or $\infty$ (or both), implying that it cannot be described over the whole space using a perturbative framework in $\zeta$. Of course, since the choice of origin is arbitrary, the divergence can happen at a finite $\chi$, for which the sum over $\ell$ will not converge.

In the current paper, we study the observational signatures of the zero modes in the CMB temperature and polarization sky. In the absence of a non-perturbative model for the zero modes, we shall assume that they remain perturbative (i.e. $\zeta({\bf x}) \ll 1$) within our observable horizon, and stick with a phenomenological description. In Sec. \ref{sec:linear} and ~\ref{sec:projection}, we summarize the linear growth history for the zero modes and outline their translation properties. Sec. \ref{sec:cmb} provides an analytic derivation of CMB anisotropy temperature and polarization power spectra for generic zero modes.  Sec. \ref{sec:constraints} forecasts the observational limits on two concrete realizations of the zero modes, namely white noise on a ``holographic screen'' and a primordial monopole. Finally, Sec. \ref{sec:conclude} concludes the paper and provides a prospectus for future lines of inquiry.

\section{Linear Growth of Zero Modes}\label{sec:linear}

We start with a self-contained summary of linear perturbation theory for zero modes, which closely follows the treatment presented in \cite{Dodelson:2003ft,Hu:1994jd} in the long wavelength limit. 
We choose to work in Newtonian gauge with metric (where we have assumed adiabatic modes and no anisotropic stress, so set the two gravitational potentials equal):
\begin{equation}
ds^2=a^2(\tau)\left[-(1+2\Psi({\bf x},\tau))d\tau^2+(1-2\Psi({\bf x},\tau) ) \left( d\chi^2 + \chi^2 d\Omega_2^2 \right) \right],
\end{equation}
where 
\begin{equation}
\zeta({\bf x}) =\Psi({\bf x},\tau) -\frac{\partial_\tau\Psi({\bf x},\tau)+(\partial_\tau\ln a)\Psi({\bf x},\tau)}{\partial_\tau\ln(-\partial_\tau a^{-1})},\label{zeta_def}
\end{equation}
 determines the curvature perturbation (or Bardeen variable) in Eq. \eqref{zeta} on superhorizon scales. 
We can split $\Psi$ into zero mode and non-zero mode contributions:
\begin{equation}
\Psi = \Psi_0 + \Psi_{\neq 0}
\end{equation}
Cosmological zero modes are defined as components of the potential $\Psi$ that satisfy the Laplace equation:
\begin{equation}
\nabla^2 \Psi_{0} = 0
\end{equation}
For all modes, the equations of motion are given by
\begin{equation}
-\nabla^2 \Psi + 3 \frac{\partial_\tau a}{a} \left[ \frac{\partial_\tau a}{a} \Psi + \partial_\tau \Psi \right] = - 4 \pi G_N a^2 \bar{\rho} \delta 
\end{equation}
\begin{equation}
\partial_\tau \delta + \nabla \cdot \vec{v} = - 3 \partial_\tau \Psi
\end{equation}
and
\begin{equation}
\partial_\tau \vec{v} + \frac{\partial_\tau a}{a} \vec{v} = - \nabla \Psi
\end{equation}
For the zero modes, Einstein's equation yields:
\begin{equation}
 3 \frac{\partial_\tau a}{a} \left[ \frac{\partial_\tau a}{a} \Psi_0 + \partial_\tau \Psi_0 \right] = - 4 \pi G_N a^2 \bar{\rho} \delta_0 
\end{equation}
This implies we can write 
\begin{equation}
\Psi_0 (\tau) = D_{\Psi}(\tau) \Psi_0 (\tau=\tau_i)
\end{equation}
From the velocity equation, we have
\begin{equation}
\vec{v}_0 (\tau) = - D_{v}(\tau) \nabla \Psi_0 (\tau=\tau_i)
\end{equation}
which implies
\begin{equation}
\nabla \cdot \vec{v}_0 = 0
\end{equation}
yelding equations for the density and velocity
\begin{equation}
\partial_\tau \delta_0 = - 3 \partial_\tau \Psi_0
\end{equation}
\begin{equation}
\partial_\tau \vec{v}_0 + \frac{\partial_\tau a}{a} \vec{v}_0 = - \nabla \Psi_0
\end{equation}
The equations for the time-dependence of the zero-modes are identical to those for modes with $k \rightarrow 0$. In $\Lambda$CDM, the growth functions are:
\bw
\be
\label{eqn:PhiSH}
D_\Psi(a)\equiv \frac{\Psi_{0}({\bf x}, \tau)}{\Psi_{0}({\bf x}, 0)}=\frac{16\sqrt{1+y}+9y^3+2y^2-8y-16}{10y^3} \left[ \frac{5}{2}\Omega_m \frac{E(a)}{a}\int_0^a
\frac{da}{E^3(a) \ a^3} \right] ,
\ee
\ew
where $E(a)=\sqrt{\Omega_ma^{-3}+\Omega_\Lambda}$ is the normalized Hubble parameter and $y=a/a_{eq}$.
\be
\label{eqn:Dv}
D_v(a)\equiv \frac{2a^2H(a)}{H^2_0\Omega_m} \frac{y}{4+3y}
\left[D_\Psi+\frac{dD_\Psi}{d\ln a}\right].
\ee
Note that, using Eq. \eqref{zeta_def}, we can see that $\zeta({\bf x})= \frac{3}{2} \Psi({\bf x}, 0)$ deep in the radiation era. 

\section{Projecting and translating zero modes}\label{sec:projection}

As we discussed in Sec. \ref{sec:intro}, the general solution for zero modes at a fixed time, specified by the scale factor $a(\tau)$, and position $\chi {\bf \hat{n}}$ can be written in terms of regular and irregular solid harmonics centered on a fiducial origin as:
\begin{eqnarray}\label{eq:origpsi}
\Psi^R (\chi {\bf \hat{n}},a) &=& \sum_{\ell m} \Psi^R_{\ell, m} (a) \ \frac{\chi^\ell}{\chi_{\rm H}(a_0)^\ell} \ Y_{\ell m}({\bf \hat{n}}) \\
\Psi^I (\chi {\bf \hat{n}},a) &=& \sum_{\ell m} \Psi^I_{\ell, m} (a) \ \frac{\chi_{\rm H}(a_0)^{\ell+1}}{\chi^{\ell+1}} \ Y_{\ell m}({\bf \hat{n}})
\end{eqnarray}
We have chosen to normalize distances to $\chi_{\rm H}(a_0)$, the comoving distance from our position to our horizon. At an arbitrary time we have:
\begin{equation}
\chi_{\rm H} (a) =  \int_{0}^{a} \ \frac{da}{H(a)a^2}.
\end{equation}
With these conventions, $\Psi^R_{\ell, m} (0)$ and $\Psi^I_{\ell, m} (0)$ are the spherical harmonic coefficients of the regular and irregular zero modes projected onto our horizon, respectively. Specifying the zero modes on the horizon is sufficient to reconstruct the entire solution both inside and outside the horizon at $a=\tau=0$; the linear evolution equations from Sec~\ref{sec:linear} can then be used to find the solution for all times within the horizon volume.

\begin{figure}
	\includegraphics[width=12cm]{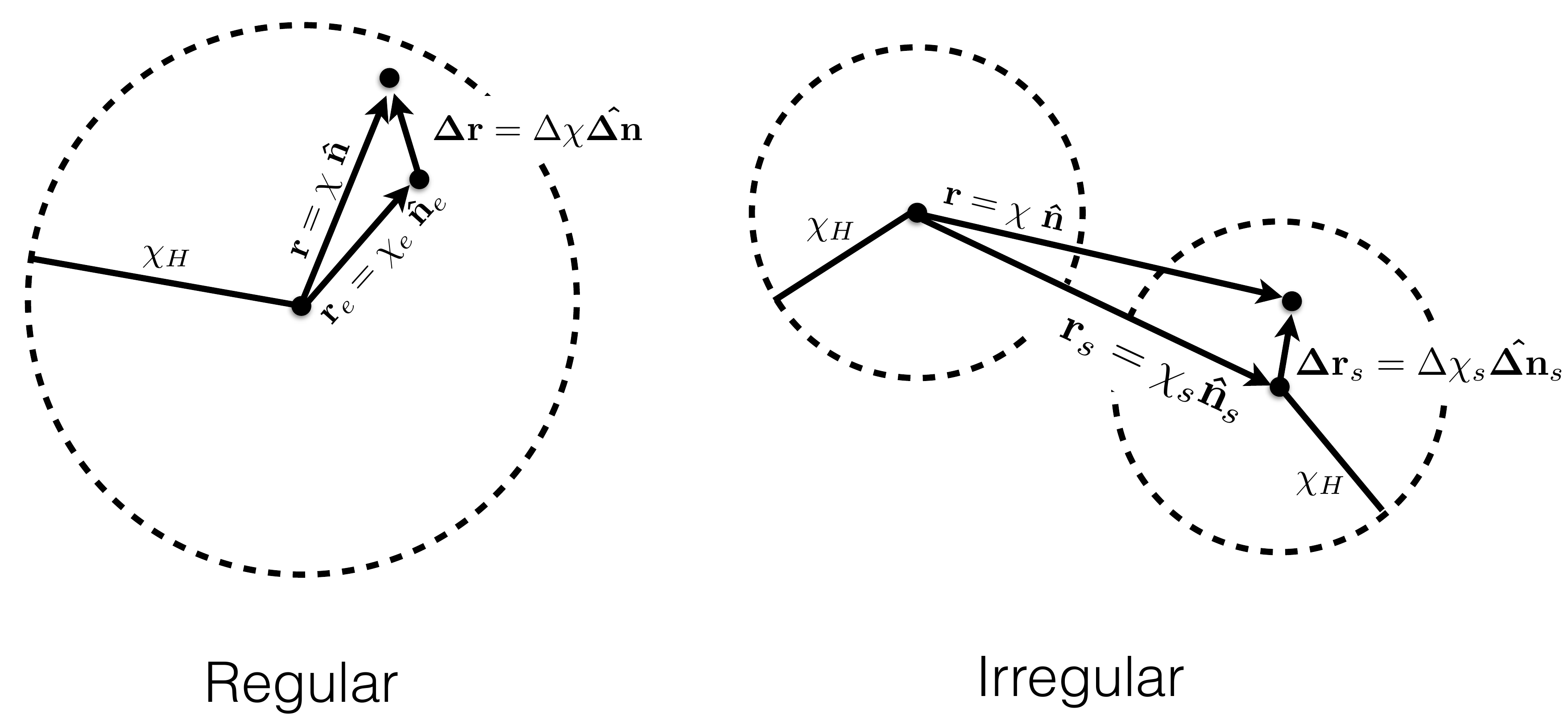}
	\caption{The geometry for regular (left) and irregular (right) zero modes. For regular zero modes, we write the position ${\bf r} = \chi {\bf \hat{n}} = {\bf r}_e + {\bf \Delta r}$ as the sum of a point on the past light cone emanating from the origin ${\bf r}_e = \chi_e {\bf \hat{n}}_e$ and distance ${\bf \Delta r} = \Delta \chi {\bf \Delta \hat{n}}$. We locate the observer at the origin of coordinates; their horizon is at a distance $\chi_H$. For the irregular modes, the singularity is located at the origin, and we consider a translated origin at position ${\bf r}_s = \chi_s {\bf \hat{n}}_s$, and denote the distance from this remote origin by ${\bf \Delta r}_s = \Delta \chi_s {\bf \Delta \hat{n}}_s$. We locate the observer at the remote origin; their horizon is again at a distance $\chi_H$.}
	\label{fig:geometry}
\end{figure}

Note that in cartesian coordinates, the regular zero modes are polynomials of order $\ell$. For example for $\ell =2$ we have:
\begin{equation}
r^2 Y_{2, -2} \propto xy, \ \ r^2 Y_{2, -1} \propto yz, \ \ r^2 Y_{2, 0} \propto -x^2-y^2+2z^2, \ \ r^2 Y_{2, 1} \propto zx, \ \ r^2 Y_{2, 2} \propto x^2-y^2.
\end{equation}
The irregular zero modes diverge at the origin of coordinates, with e.g. the $\ell = m = 0$ component diverging as $\Psi^I \propto \chi^{-1}$ as $\chi \rightarrow 0$. For the regular harmonics, the fiducial origin of coordinates defined by Eq.~\eqref{eq:origpsi} could be chosen as our location. For the irregular harmonics, we locate the singularity outside of our horizon volume so that we can apply cosmological perturbation theory within our horizon. See Fig.~\ref{fig:geometry} for the geometry in the case of regular and irregular zero modes.

Given the solution for the zero modes at a fixed time, we can use the translation properties of solid harmonics to find the zero modes as observed from a remote position. This will be necessary for determining the CMB temperature and polarization anisotropies induced by zero modes. As shown in Fig.~\ref{fig:geometry}, for regular zero modes we locate the remote origin within the horizon at ${\bf r}_e = \chi_e \hat{n}_e$ and specify the spatial distance from the remote origin by ${\bf \Delta r} = \Delta \chi {\bf \Delta \hat{n}}$. For irregular zero modes, we locate the singular origin outside the horizon at ${\bf r}_s = \chi_s \hat{n}_s$ and specify the spatial distance from the remote origin by ${\bf \Delta r}_s = \Delta \chi_s {\bf \Delta \hat{n}}_s$. Performing an expansion in regular solid harmonics around the remote origin, we have:
\begin{eqnarray}\label{eq:translatedpsis}
\tilde{\Psi}^R &=& \sum_{\ell m} \tilde{\Psi}^R_{\ell, m} \ \frac{\Delta \chi^\ell}{\chi_{\rm H}(a_e)^\ell} \ Y_{\ell m}({\bf \hat{\Delta n}}) \\
\tilde{\Psi}^I &=& \sum_{\ell m} \tilde{\Psi}^I_{\ell, m} \  \frac{\Delta \chi_s^\ell}{\chi_{\rm H}(a_e)^\ell} \ Y_{\ell m}({\bf \hat{\Delta n}}_s)
\end{eqnarray}
where
\begin{equation}
\chi_e = \int_{a_e}^{1} \ \frac{da}{H(a)a^2}, \ \ \ \Delta\chi(a)= \int_{a}^{a_e} \frac{da}{H(a)a^2}
\end{equation}

The solid harmonics are translated using the relations (see e.g.~\cite{citeulike:10828776}):
\begin{eqnarray}\label{eq:translation}
\chi^\ell Y_{\ell,m} (\hat{n}) &=& \sum_{\ell' m'} C^R(\ell,m,\ell',m') \ \Delta \chi^{\ell'} Y_{\ell' m'} ({\bf \hat{\Delta n}}) \ \chi_e^{\ell-\ell'} Y_{\ell-\ell',  m-m'} ({\bf \hat{n}}_e) \\
\frac{1}{\chi^{\ell+1}} Y_{\ell,m} (\hat{n}) &=& \sum_{\ell' m'} C^I(\ell,m,\ell',m') \ \Delta\chi_s^{\ell'} Y^*_{\ell' m'}  ({\bf \hat{\Delta n}}_s )  \ \frac{1}{\chi_s^{\ell+\ell' + 1}} Y_{\ell+\ell',  m+m'}({\bf \hat{n}}_s)
\end{eqnarray}
where
\begin{eqnarray}
C^R(\ell,m,\ell',m') &=& \sqrt{\frac{4 \pi (2\ell+1)}{(2\ell - 2 \ell'+1) (2 \ell'+1)}} \sqrt{\frac{(\ell-m)! (\ell+m)!}{(\ell'-m')! (\ell'+m')!}} \frac{1}{\sqrt{(\ell-\ell'-m+m')! (\ell-\ell'+m-m')!}} \\
C^I(\ell,m,\ell',m') &=& (-1)^{\ell'}  \sqrt{\frac{4 \pi (2\ell+1)}{(2\ell + 2 \ell'+1) (2 \ell'+1)}}  \frac{\sqrt{(\ell+\ell'-m-m')! (\ell+\ell'+m+m')!}}{ \sqrt{(\ell+m)! (\ell-m)!(\ell'+m')!(\ell'-m')!}}
\end{eqnarray}
Using these relations, the expansion coefficients in Eq.~\eqref{eq:origpsi} and Eq.~\eqref{eq:translatedpsis} are related by
\begin{eqnarray}\label{eq:translatedcoeffs}
\tilde{\Psi}^R_{\ell, m} &=& \sum_{\ell'' m''} \Psi^R_{\ell'', m''} \ \frac{\chi_{\rm H}(a_e)^{\ell}}{\chi_e^\ell}  \frac{\chi_e^{\ell''}}{\chi_{\rm H}(a_0)^{\ell''}} \ C^R(\ell'',m'',\ell,m) Y_{\ell''-\ell,  m''-m} ({\bf \hat{n}}_e) \\
\tilde{\Psi}^I_{\ell, m} &=& \sum_{\ell'' m''} \Psi^I_{\ell'', m''} \frac{\chi_{\rm H}(a_0)^{\ell+\ell''+1}}{\chi_s^{\ell+\ell''+1}} C^I(\ell'',m'',\ell,m) Y_{\ell+\ell'',  m+m''} ({\bf \hat{n}}_s)
\end{eqnarray}

As a quick check of the result above, for the regular zero modes when $\chi_e \rightarrow 0$, $\chi \rightarrow \Delta \chi$, and $\chi_{\rm H}(a_e) \rightarrow \chi_{\rm H}(a_0)$ we should obtain $\tilde{\Psi}^R_{\ell, m} = \Psi^R_{\ell, m}$. For the regular solution, the only term that survives this limit in the sum is the term for $\ell'' = \ell$. Here, we have $C^R(\ell,m,\ell,m) = 2 \sqrt{\pi}$ and the spherical harmonic factor evaluates to $Y_{\ell''-\ell,  m''-m} ({\bf \hat{n}}_e) = Y_{0, 0} ({\bf \hat{n}}_e) = 1/(2\sqrt{\pi})$, and so we do indeed find that $\tilde{\Psi}^R_{\ell, m} = \Psi^R_{\ell, m}$ in the appropriate limit. Because we have expanded the irregular zero modes in terms of regular modes at the remote origin, it is not possible to take an analogous limit in this case.

\section{CMB signatures of zero modes}\label{sec:cmb}

The CMB temperature anisotropy as viewed from ${\bf r}_e = \chi_e {\bf \hat{n}}_e $ receives three contributions (see e.g., \cite{Dodelson:2003ft}):
\ba \label{eq:electron_CMB}
\Theta({\bf \hat{n}}_e, \chi_e, {\bf \hat{\Delta n}}) &=& \Theta _{\rm SW} ({\bf \hat{n}}_e, \chi_e, {\bf \hat{\Delta n}}) + \Theta_{\rm Doppler} ({\bf \hat{n}}_e, \chi_e, {\bf \hat{\Delta n}}) + \Theta _{\rm ISW}({\bf \hat{n}}_e, \chi_e, {\bf \hat{\Delta n}}).
\ea
The various components are given by:
\be \label{eq:thetaSW}
\Theta _{\rm SW}= \int da \ g(a) \left( 2 D_\Psi(a) -\frac{3}{2} \right) \tilde{\Psi}^{R,I} (\Delta \chi (a) {\bf \hat{\Delta n}}),
\ee
\be\label{eq:thetaDopp}
 \Theta_{\rm Doppler} =  {\bf \hat{\Delta n}}\cdot
[{\bf v}({\bf r}_e,\chi_e)  -  \int da \ g(a) \ {\bf v}({\bf \Delta r} (a))],
\ee
\be
\label{eqn:ISW}
\Theta _{\rm ISW} =2\int_{0}^{a_e} da \ e^{-\tau(a)} \ \frac{dD_\Psi}{da} \tilde{\Psi}^{R,I} (\Delta \chi (a) {\bf \hat{\Delta n}}),
\ee
where SW and ISW stand for Sachs-Wolfe and Integrated Sachs-Wolfe effects and we have defined the visibility function
\begin{equation}
g(a) = \frac{d \tau (a)}{d a} \ e^{-\tau (a)}, \ \ \ \int d a \ g(a) = 1.
\end{equation}
We obtain the recombination history using the RECFAST++ code~\cite{2010MNRAS.403..439R,Chluba:2010ca,doi:10.1111/j.1365-2966.2009.15957.x,doi:10.1111/j.1365-2966.2010.16940.x,Seager:1999bc} and model the reionization ionization fraction by:
\begin{equation}
X_{\rm reion}(z) = \frac{1}{2} \left[ 1 - \tanh \left( \frac{(z-z_{\rm reion})}{\Delta z} \right) \right],
\end{equation}
where we choose $z_r=8.4$ and $\Delta z=0.5$. 

\subsection{Temperature}

\begin{figure}[htbp]
	\begin{center}
	\includegraphics[width=0.45\textwidth]{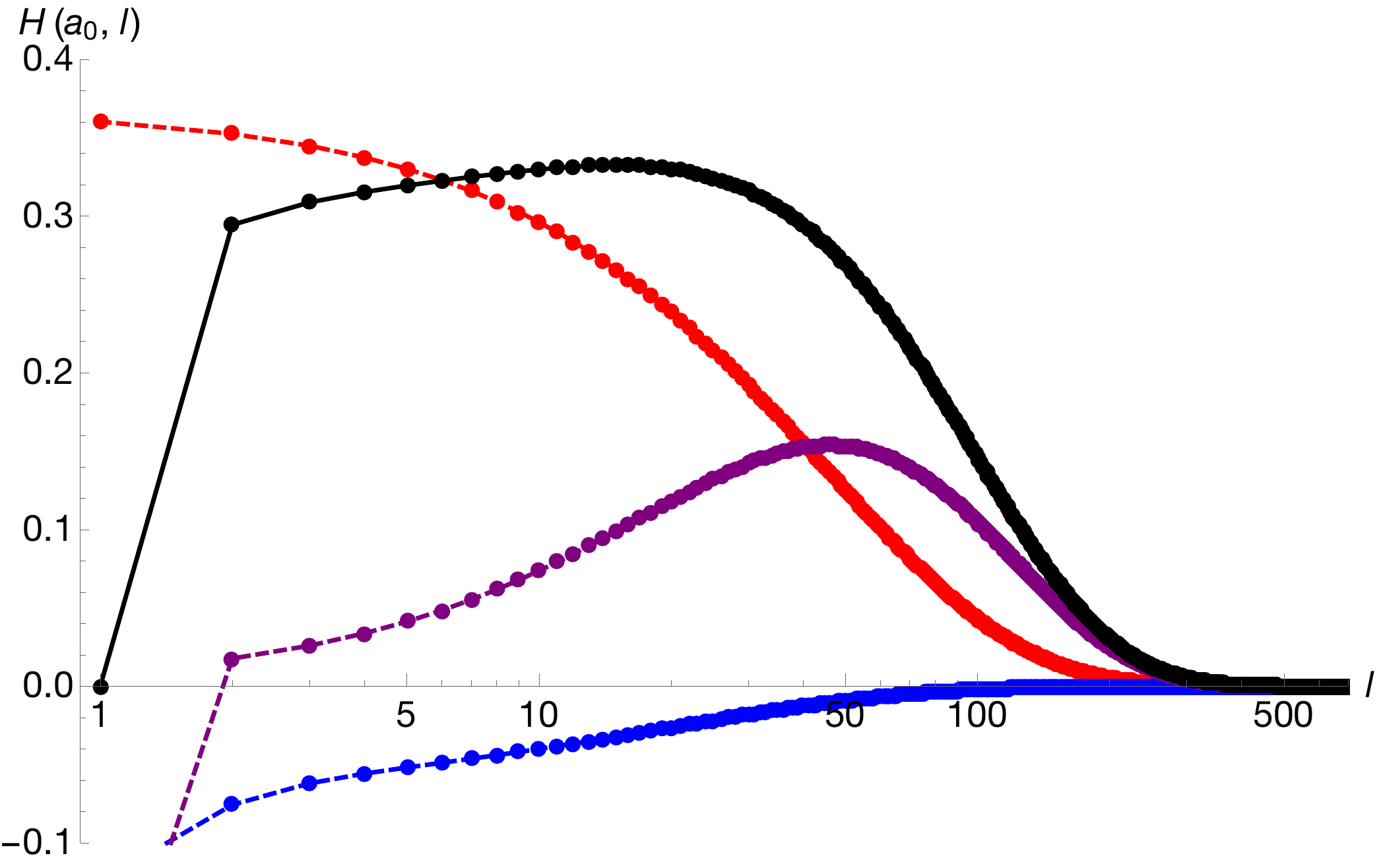}
	\includegraphics[width=0.45\textwidth]{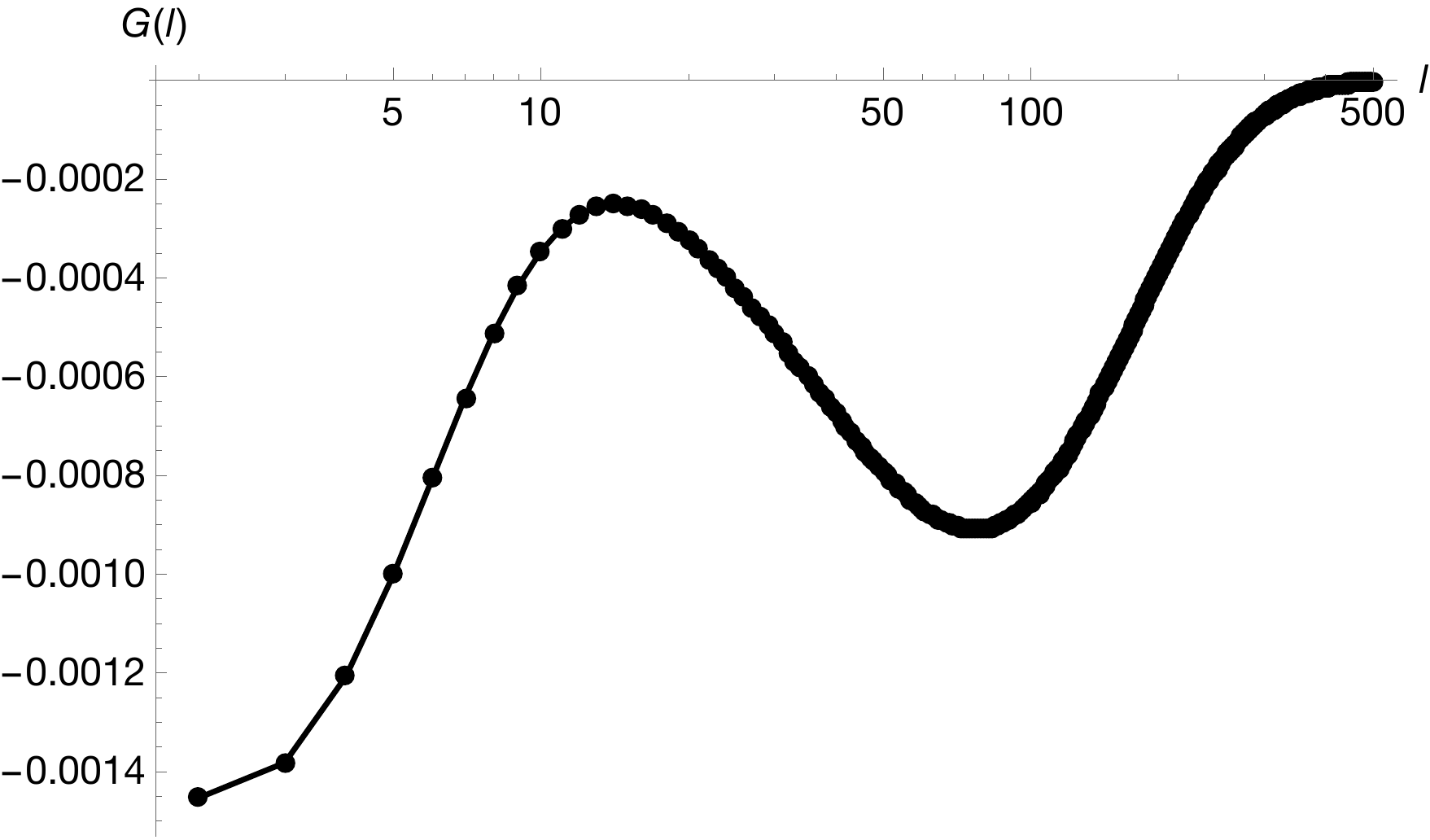}
	\caption{The projection kernels from the regular zero mode multipoles to the CMB temperature (left) and E-mode polarization (right) multipoles in $\Lambda$CDM cosmology. Solid black is the total function, red dashed the SW, purple the doppler, and blue the ISW contributions.}
	\label{fig:kernel}
	\end{center}
\end{figure}

Focusing first on the regular modes, we obtain the CMB temperature anisotropies observed at our position by setting $\chi_e=0$ in Eq.~\eqref{eq:electron_CMB} and taking the spherical harmonic transform. This yields:
\begin{equation}
\Theta^R_{\ell m} = H^R(a_0,\ell) \Psi_{\ell, m}^R
\end{equation}
where
\bw
\begin{equation}\label{eq:Hr}
H^R(a,\ell) = \int_0^a da \left(\frac{\Delta \chi (a)}{\chi_H(a_0)}\right)^\ell \left[ g(a) \left( 2D_\Psi(a) -\frac{3}{2} \right) 
+ 2 e^{-\tau(a)}\frac{dD_\Psi}{da} 
 + \ell g(a) \frac{D_v(\chi_{\rm dec}) - D_v(0)\delta_{\ell, 1} }{\Delta \chi (a) }  \right].
\end{equation}
We show the kernel Eq.~\eqref{eq:Hr} in the left panel of Fig.~\ref{fig:kernel}. For $\ell = 1$, the zero modes correspond to a pure gradient in the Newtonian potential, which can be absorbed by a gauge transformation (see e.g.~\cite{Erickcek:2008jp,Terrana2016}) and therefore should not lead to any observational signatures. There is indeed no temperature dipole induced by zero modes, as Eq.~\eqref{eq:Hr} vanishes for $\ell = 1$. Examining the relative contributions from the SW, ISW, and Doppler terms we see that SW dominates on the largest angular scales, but the Doppler contribution dominates at a relatively low $\ell$, near $\ell \agt 50$. This can be traced back to the fact that zero modes for large $\ell$ correspond to high order polynomials, which have large gradients, and therefore source large velocities, near the recombination surface. 

\ew

\subsection{Polarization}
The stokes parameters describing CMB polarization anisotropies are given by:
\ba
(Q \pm iU) ({\bf \hat{n}}_e) = - \frac{\sqrt{6}}{10}  \int da_e \ g(a_e)   \sum_{m=-2}^{2} \Theta_{2,m} ({\bf r}_e(a_e)) \ \left._{\pm 2}Y_{2 m}\right. ({\bf \hat{n}}_e) 
\label{eq:effective-quadrupole-to-QiU}
\ea
where $\Theta_{2,m} ({\bf r}_e(a_e))$ is the locally observed CMB temperature quadrupole at the position ${\bf r}_e$ along our past light cone. The stokes parameters are decomposed into spin-2 harmonics as:
\begin{equation}
(Q \pm iU)({\bf \hat{n}}_e) = \sum_{\ell,m}  a_{\pm2,\ell m}   \left._{\pm 2}Y_{\ell m}\right.({\bf \hat{n}}_e)
\end{equation}
where the coefficients satisfy $a^*_{-2,\ell m} = a_{2,\ell -m}$. The scalar E mode multipoles are defined as 
\begin{equation}
a_{E,\ell m} = -\frac{1}{2} (a_{2,\ell m} + a_{-2,\ell m})
\end{equation}
with
\begin{equation}
E({\bf \hat{n}}) = \sum_{\ell,m} a_{E,\ell m} Y_{\ell m}({\bf \hat{n}})
\end{equation}
B-mode multipoles are defined as
\begin{equation}
a_{B,\ell m} = -\frac{1}{2i} (a_{2,\ell m} - a_{-2,\ell m})
\end{equation}

To compute the polarization signal, we need to find the temperature quadrupole observed at the position of each scatterer. Re-centering the coordinate system around an electron at $\chi_e {\bf \hat{n}}_e$, and using Eqs.~\eqref{eq:translatedcoeffs} we find 
\bw
\begin{eqnarray}
\Theta_{2,m} ({\bf \hat{n}}_e, \chi_e) &=& H^R(a_e,\ell=2) \tilde{\Psi}_{2, m}^R \\
&=&  \sum_{\ell' m'} \Psi^R_{\ell', m'} \ H^R(a_e,\ell=2)\frac{\chi_{\rm H}(a_e)^{2}}{\chi_e^2}  \frac{\chi_e^{\ell'}}{\chi_{\rm H}(a_0)^{\ell'}} \ C^R(\ell',m',2,m) Y_{\ell'-2,  m'-m} ({\bf \hat{n}}_e)
\end{eqnarray}

The spin-2 multipoles of $(Q \pm iU)({\bf \hat{n}}_e)$ are:
\begin{eqnarray}
a_{\pm2,\ell m} &=& \int (Q \pm iU)({\bf \hat{n}}_e)  \left._{\pm 2}Y_{\ell m}\right.^*({\bf \hat{n}}_e) d^2{\bf \hat{n}}_e \\
&=& - \frac{\sqrt{6}}{10}  \int da \ g(a) \sum_{m''=-2}^{2} \int  \Theta_{2,m''} ({\bf \hat{n}}_e, \chi_e) \ \left._{\pm 2}Y_{2 m''}\right. ({\bf \hat{n}}_e)    \left._{\pm 2}Y_{\ell m}\right.^*({\bf \hat{n}}_e) d^2{\bf \hat{n}}_e \\
&=& - \frac{\sqrt{6}}{10}  \int da \ g(a)  H^R(a,\ell=2) \sum_{m''=-2}^{2} \int   \tilde{\Psi}^R_{2,m''} ({\bf \hat{n}}_e, \chi_e) \ \left._{\pm 2}Y_{2 m''}\right. ({\bf \hat{n}}_e)    \left._{\pm 2}Y_{\ell m}\right.^*({\bf \hat{n}}_e) d^2{\bf \hat{n}}_e \\
&=& - \sum_{\ell', m'} \sum_{m''=-2}^{2} \frac{\sqrt{6}}{10}  \int da \ g(a) H^R(a,\ell=2)  \frac{\chi_{\rm H}(a_e)^{2}}{\chi_e^2}  \frac{\chi_e^{\ell'}}{\chi_{\rm H}(a_0)^{\ell'}}  \Psi_{\ell', m'} C^R(\ell',m',2,m'')  \\  &\times& \int   \ \left._{\pm 2}Y_{\ell m}\right.^*({\bf \hat{n}}_e) \  \left._{\pm 2}Y_{2 m''}\right. ({\bf \hat{n}}_e) \ Y_{\ell'-2,  m'-m''} ({\bf \hat{n}}_e) d^2{\bf \hat{n}}_e
\end{eqnarray}
We can evaluate the integral over hamonics using 3j symbols, yielding:
\begin{equation}
a_{E,\ell m} = G(\ell) \Psi_{\ell, m}
\end{equation}
where we have defined
\begin{eqnarray}
G(\ell) &\equiv& - \sum_{m''=-2}^{2} \frac{\sqrt{6}}{10}  \int da \ g(a) H^R(a,2)  \frac{\chi_{\rm H}(a)^{2}}{\chi_e^2}  \frac{\chi_e^{\ell}}{\chi_{\rm H}(a_0)^{\ell}}   C^R(\ell,m,2,m'') \\ &\times& (-1)^m \sqrt{\frac{5(2\ell+1)(2\ell-3)}{4\pi}} \left(\begin{array}{ccc} \ell & 2 & \ell -2 \\ -m &m''  & m-m'' \end{array} \right)  \left(\begin{array}{ccc} \ell & 2 & \ell-2 \\ 2 &- 2 & 0 \end{array} \right),
\end{eqnarray}
Note that while $m$ appears in this formula, the result of evaluating the full expression is independent of $m$. We show $G(\ell)$ in the right panel of Fig.~\ref{fig:kernel}. As expected, the polarization anisotropies are smaller than temperature anisotropies. The double-peak structure of $G(\ell)$ is due to the separate contributions associated with recombination (high-$\ell$) and reionization (low-$\ell$). Finally, note that the temperature and $E$-mode polarization anisotropies take opposite sign, and therefore the contribution to the TE correlation function from zero modes is negative definite.
 
\ew

\section{Constraining Zero Modes on the ``holographic screen'' }\label{sec:constraints}

The uniqueness theorem for the solutions of the Laplace equation implies that specifying the boundary conditions on the cosmological horizon completely fixes our observable zero modes. This can be seen explicitly in Eq. \eqref{eq:origpsi} by setting $\chi=\chi_H$:
\begin{equation}
\Psi^R (\chi_H {\bf \hat{n}},0) = \frac{2}{3} \zeta(\chi_H {\bf \hat{n}})=  \sum_{\ell m} \Psi^R_{\ell, m} (0) Y_{\ell m}({\bf \hat{n}}),
\end{equation}
i.e. decomposing the metric perturbations on the cosmological horizon into spherical harmonics specifies $ \Psi^R_{\ell, m} (0)$'s, which in turn fixes $\Psi^R$ throughout spacetime using Eqs. \eqref{eqn:PhiSH} and \eqref{eq:origpsi}. In other words, the cosmological horizon plays the role of a ``holographic screen'' for the zero modes. 

In this section, we consider two representative models for the spectrum of zero modes on this ``holographic screen'' and forecast the possible constraints in the cosmic variance limit. The zero modes on the screen can range from fully incoherent to fully coherent. For the former, we consider a white noise for the spectrum of $\Psi^R$ on the horizon, while for the latter, we consider a single ``monopole'', or irregular $\ell =0$ mode, beyond the horizon. 

\subsection{White noise on the ``holographic screen''}
\begin{figure}[htbp]
	\begin{center}
	\includegraphics[width=5cm]{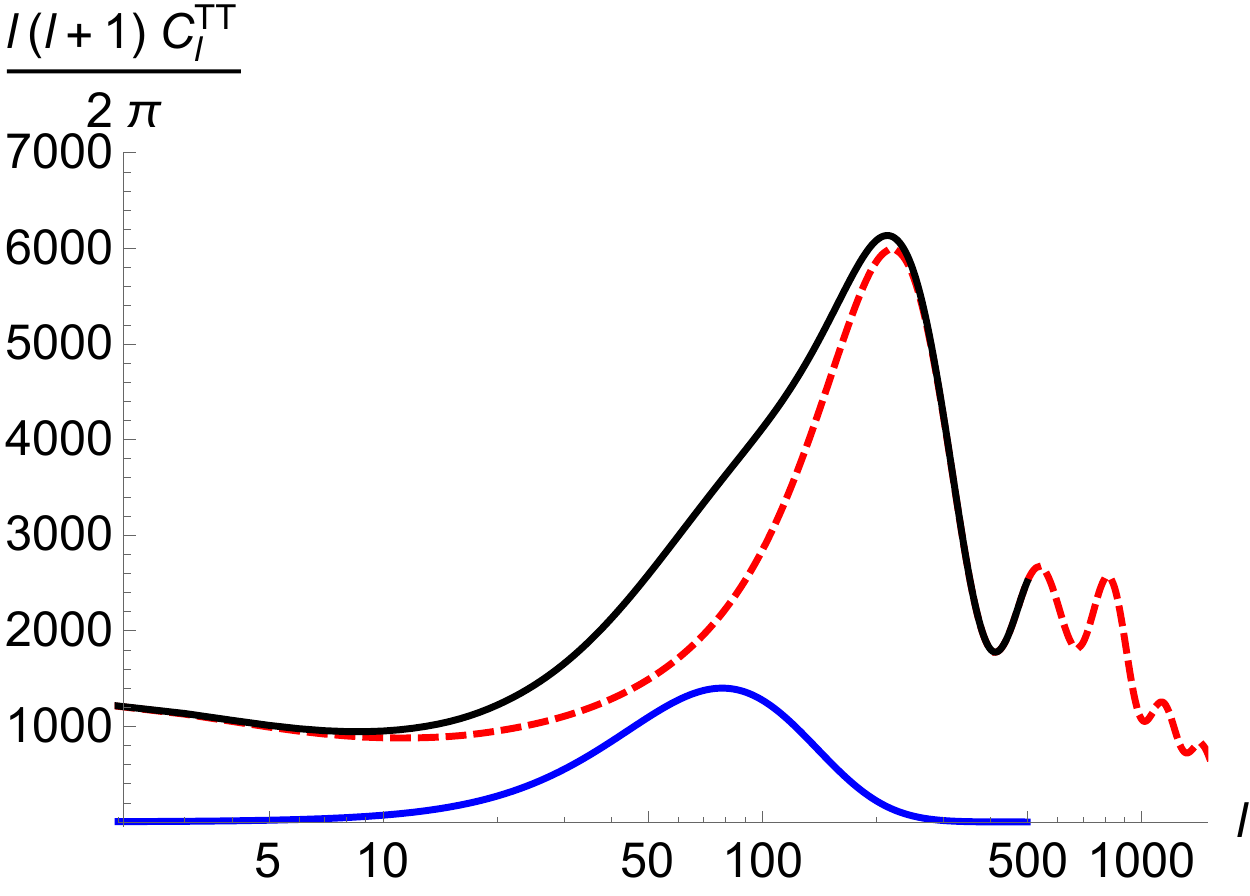}
	\includegraphics[width=5cm]{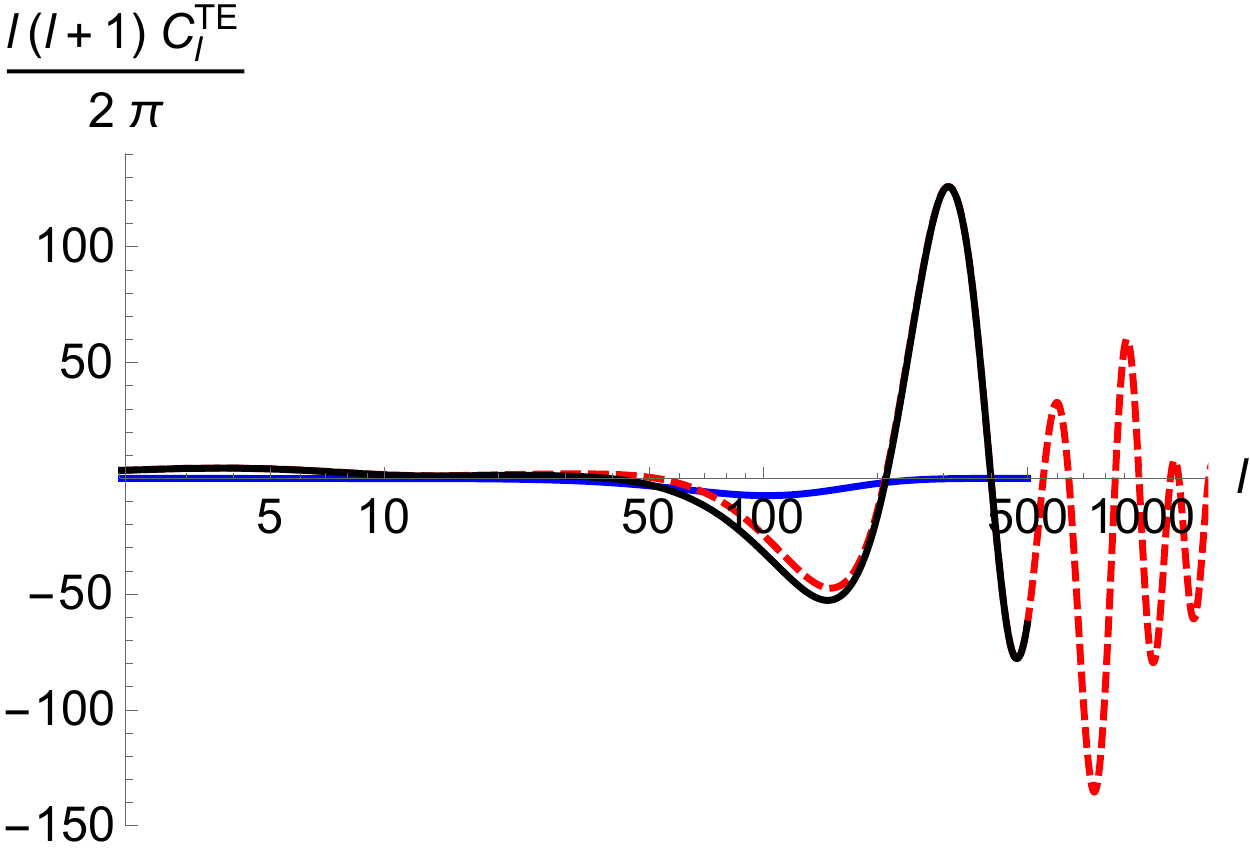}
	\includegraphics[width=5cm]{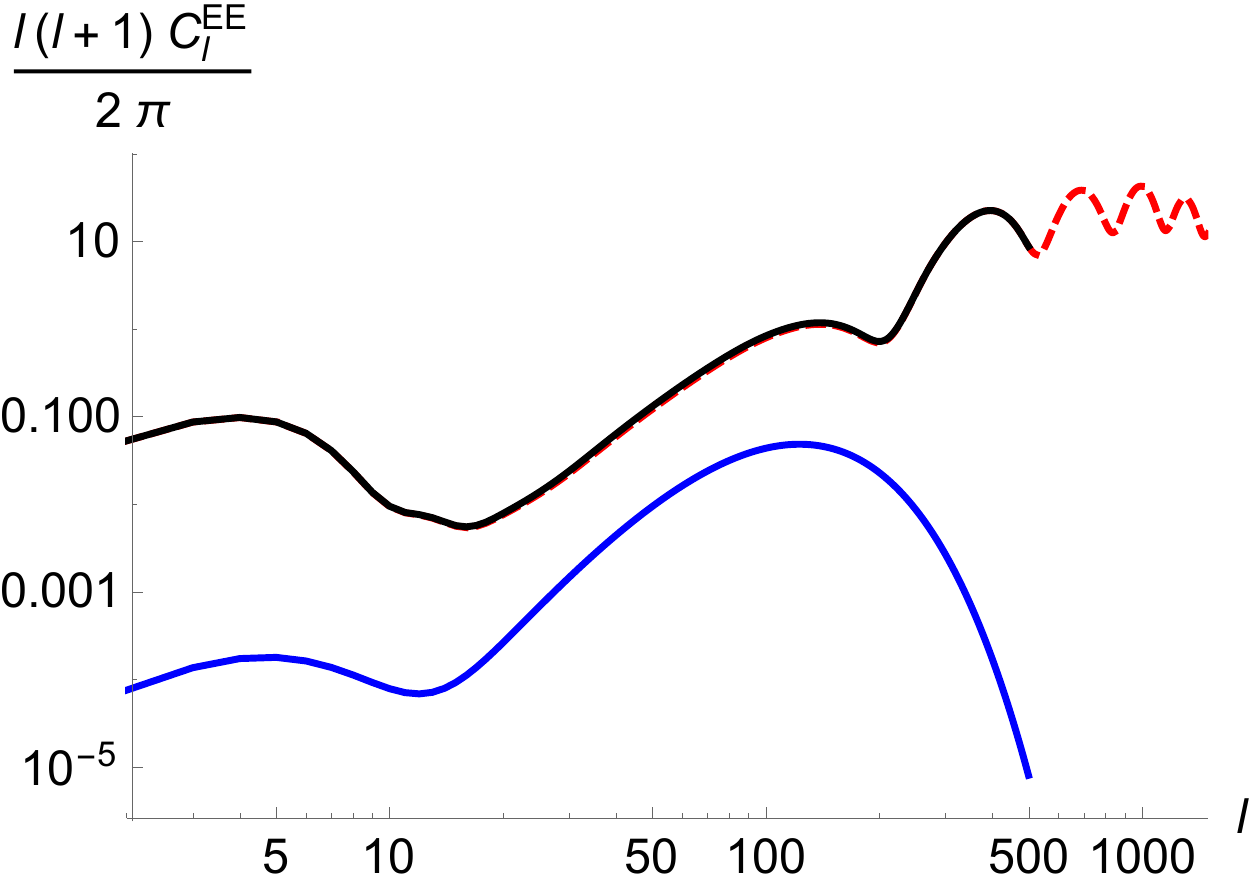}
	\caption{The TT (left), TE (center), and EE (right) power spectra for uncorrelated zero modes with $A=5 \times 10^{-12}$ (blue solid). Shown also are power spectra for $\Lambda$CDM (red dashed) and the sum of the $\Lambda$CDM power and uncorrelated zero modes (black solid).}
	\label{fig:pspec}
	\end{center}
\end{figure}

In the first scenario, we consider a completely uncorrelated set of zero modes on the horizon with amplitude $A$ described by:
\begin{equation}
\langle \Psi^R_{\ell,m} \Psi^R_{\ell',m'}\rangle = A \ \delta_{\ell \ell'} \delta_{m m'}
\end{equation}
This gives a contribution to the CMB temperature and polarization power spectra given by
\begin{equation}
{\bf C}^0_\ell \equiv \begin{pmatrix} C_\ell^{TT,0} & C_\ell^{TE,0} \\ C_\ell^{TE,0} & C_\ell^{EE,0} \end{pmatrix} = T_{\rm CMB}^2 A \begin{pmatrix} H^R(a_0,\ell)^2 & H^R(a_0,\ell) G(\ell) \\ H^R(a_0,\ell) G(\ell) & G(\ell)^2 \end{pmatrix}
\end{equation}
In Fig.~\ref{fig:pspec} we show the contribution to the temperature and polarization power spectra from zero modes for $A=5 \times 10^{-12}$ compared with the $\Lambda$CDM power spectrum. The contribution to TT is significant, while the contribution to EE is completely negligible; the cross spectrum is slightly affected. The zero modes in this scenario therefore add power to the temperature power spectrum without adding as much power to polarization as would be expected from a statistically homogeneous random field. In addition, for this model, most of the constraining power will come from intermediate multipoles $30 \alt \ell \alt 200$ where we have good temperature and polarization data from the Planck satellite.

The Fisher information matrix in this scenario is given by
\begin{equation}
F = \sum_{\ell = 2}^{\ell_{\rm max}} \frac{2 \ell+1}{2} {\rm Tr} \left[ \left({\bf C}_\ell^{-1} \cdot \partial_A {\bf C}^0_\ell  \right)^2 \right]
\end{equation}
where ${\bf C}_\ell$ is the covariance matrix in $\Lambda$CDM. Evaluating to $\ell_{\rm max} = 500$, we obtain 
\begin{equation}
\Delta A = F^{-1/2} = 9.16 \times 10^{-14}. 
\end{equation}
We can therefore define the detectable range of amplitudes as roughly $A > \Delta A$.  

\subsection{Primordial Monopole}\label{sec:monopole}
In the second scenario, we consider the first irregular zero mode with charge $Q$:
\begin{equation}
\Psi^I = Q \frac{\chi_{\rm H}(a_0)}{\chi}
\end{equation}
and translate the center at a distance $d \equiv \chi_s / \chi_{\rm H}(a_0)$ away in the ${\bf \hat{n}}_s$ direction from our location. For simplicity, we choose ${\bf \hat{n}}_s$ to be at the north pole of the celestial sphere, in which case after translating we obtain:
\begin{equation}
\tilde{\Psi}^I_{\ell 0} = \frac{2 \sqrt{\pi}}{\sqrt{2\ell+1}}  \frac{Q}{d^{\ell+1}}  \\
\end{equation}
where we restrict $d > 1$ so that the singularity is outside of our horizon. The resulting signature in temperature and polarization is azimuthally symmetric and varies primarily on large angular scales. Constraints on the 2 parameters $Q$ and $d$ are obtained from the following Fisher matrix:
\begin{equation}
F_{ij} = \frac{1}{2} \sum_{\ell=2}^{\ell_{\rm max}} {\rm Tr} \left[ {\bf C}^{-1}_\ell \left( \partial_i {\mathbf \mu_\ell} \partial_j {\bf \mu_\ell}^T + \partial_i {\bf \mu_\ell}^T \partial_j {\bf \mu_\ell} \right)  \right]
\end{equation}
where $\mu_\ell = (a_{T, \ell 0} , a_{E, \ell 0}) $ as defined above and ${\bf C}_\ell$ is the covariance matrix in $\Lambda$CDM. 

\begin{figure}[htbp]
	\begin{center}
	\includegraphics[width=10cm]{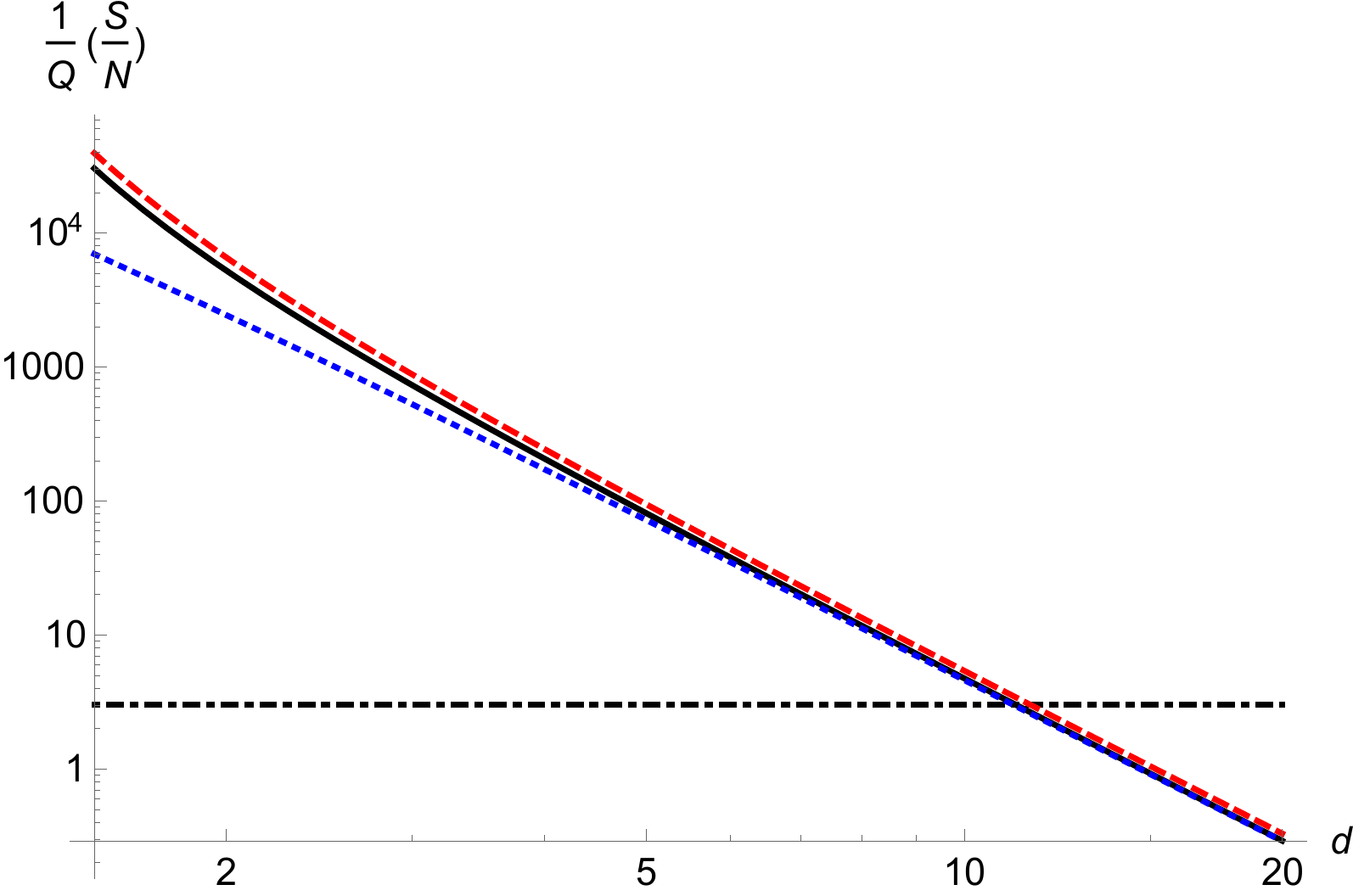}
	\caption{The signal to noise, normalized by the charge $Q$, for the primordial monopole scenario. Curves are shown for $\Lambda$CDM with $\ell_{\rm max} = 200$ (black solid), $\Lambda$CDM with $\ell_{\rm max} = 3$ (blue dotted), and the Planck measured temperature power spectrum (red dashed). The black dot-dashed line corresponds to a signal to noise of three.}
	\label{fig:BHSNplot}
	\end{center}
\end{figure}

In Fig.~\ref{fig:BHSNplot}, we show the signal to noise for a measurement of $d$ defined as $(S/N) \equiv d / \Delta d = d / \sqrt{F_{dd}^{-1}}$; this scales linearly with the charge $Q$. To get an idea for which angular scales contribute, we show curves for $\ell_{\rm max} = 3$ (blue dotted) and $\ell_{\rm max}=200$ (black solid) using the temperature and poalrization power spectrum of $\Lambda$CDM. If the monopole is not located too far away, we see that a range of multipoles contribute to the signal. However, at distances of order $\agt 5 \chi_H(a_0)$, nearly all of the constraining power comes from the quadrupole and octupole. The observed CMB at low-$\ell$ has significantly less power than predicted in $\Lambda$CDM. To investigate the effect of this on our ability to constrain zero modes, we use the CMB temperature power spectrum observed by Planck~\cite{Aghanim:2015xee} to compute the Fisher matrix (setting the low-$\ell$ polarization power $\ell<30$ to zero) and obtain the red dashed curve in Fig.~\ref{fig:BHSNplot}. Due to the lower power, the costraints on distance improve slightly. Also, due to the limited signal in the template, the low-$\ell$ polarization does not contribute significantly to the constraints.  For $\chi_s \gtrsim 5 \chi_H$ the forecast takes the simple form:
\begin{equation}
S/N \simeq 3Q \left(\chi_s \over 11.6 \chi_H\right)^{-4} =   3Q \left(\chi_s \over 160~ {\rm Gpc}\right)^{-4}, 
\end{equation}
implying that a ``monopole'' with $Q \sim 1$ can be detected at $>3\sigma$ up to a distance of $\sim 12 \times$ the cosmological horizon radius, or $\sim$160 Gpc.  

\section{Conclusion, Discussion, Future Prospects}\label{sec:conclude}

In summary, we have introduced {\it cosmological zero modes} as a target of opportunity for current and future cosmological surveys to unearth fundamentally new information about non-perturbative processes in the early universe. These zero modes are fully fixed by the boundary conditions on a ``holographic screen'' on our cosmological horizon. We have computed a transfer function that maps the zero-mode multipoles on the screen to the CMB temperature and polarization multipoles. We have also provided forecasts for observational constraints on fully incoherent (a white noise) or coherent (a primordial monopole) spectrum of zero modes on the holographic screen. 

Here, we shall outline some of the future steps and/or questions in regard to the study of zero modes:
\begin{itemize}

\item On the theoretical front, while we presented a linear definition of zero modes, one may wonder whether they can be defined non-linearly. A natural covariant definition could be:
\begin{equation}
R^{(3)} = 0,{~{\rm and}~~} C^j_i \equiv \epsilon^{klj} \nabla_k \left(R_{li} - \frac{1}{4} R g_{li}\right) =0,
\end{equation} 
on constant temperature hypersurfaces. Here, the vanishing of the Cotton-York tensor $C^j_i$ ensures that the 3-metric is conformally flat (reducing to scalar modes in the linear regime), while the vanishing of the Ricci scalar, $R^{(3)}$, imposes the zero mode condition. A more difficult question will be the possible modulation of non-zero modes by zero modes, which will be expected beyond the linear perturbation theory. 

\item One promising framework for studying zero modes is cosmic inflation. In the case where the pre-inflationary Universe contains singular regions interspersed among proto-inflationary patches, it is likely that a local description of the perturbed cosmological spacetime will include zero modes. This could be assessed using numerical relativity simulations of inhomogeneous inflation, e.g.,~\cite{East:2015ggf,Clough:2016ymm}. Furthermore, the so-called super-curvature modes in an open universe \cite{Lyth:1995cw} behave like regular zero modes (\ref{eq:origpsi}) on scales much smaller than the radius of curvature. Therefore, in the appropriate limit, open inflation~\cite{Gott:1982zf,Bucher:1994gb} provides an example of the importance of zero modes.  

\item The observed CMB temperature anisotropies exhibit a lack of power on large angular scales, low power in the quadrupole, and anomalous alignment between the octupole and quadrupole moments when compared with the expectation from $\Lambda$CDM (see e.g.~\cite{Schwarz:2015cma} for a review). Cosmological zero modes may play a role in resolving these anomalies. For example, we have seen in Sec.~\ref{sec:monopole} that a monopole configuration primarily gives rise to a temperature quadrupole and octupole, which are aligned along the direction to the monopole. To accommodate the lack of power on large scales, one would additionally have to invoke a suppression of power in adiabatic modes. This may be expected in a scenario with just enough inflation, which may also provide an explanation for having only one observable monopole configuration and not zero or many (which would remove the alignment between the quadrupole and octupole). Extending Ref.~\cite{Liddle:2013czu}, there are also potentially interesting consequences of zero modes for other CMB anomalies such as the hemispherical power asymmetry. 

\item Progress could be made assessing scenarios involving cosmological zero modes through the analysis of existing CMB temperature and polarization data from the Planck satellite. We defer this exercise to future work, although our forecasts in Sec.~\ref{sec:constraints} provide a preliminary assessment of the level at which zero modes could be detected or constrained. 

\item The three dimensional structure of cosmological zero modes is very different from the adiabatic fluctuations in $\Lambda$CDM. Therefore, measures of homogeneity on ultra large scales could greatly improve the prospects for observing or constraining cosmological zero modes. Measurements of the remote CMB dipole and quadrupole using kinetic Sunyaev Zel'dovich tomography~\cite{Zhang10d,Zhang:2015uta,Terrana2016} and polarized Sunyaev Zel'dovich tomography~\cite{Kamionkowski1997,Deutsch:2017cja,Deutsch:2017ybc} are well suited for constraining homogeneity on ultra large scales, making these ideal observational probes for cosmological zero modes.

\item Let us end by speculating on the holographic nature of zero modes and whether this can be related to modern developments in holography as related to quantum gravity. One may notice that, had we lived in an open universe (i.e. Euclidean AdS$_3$) , and let the ``holographic screen'' go to $\chi \rightarrow \infty$, we may expect a cosmological description in terms of a 2d conformal field theory \cite{Freivogel:2006xu}. This invites a more rigorous study into the holographic nature of zero (and non-zero) modes.   

\end{itemize}

\begin{acknowledgments}
This work has been partially supported by Perimeter Institute for Theoretical Physics (PI). Research at PI is supported by the 
Government of Canada through the Department of Innovation, Science and Economic Development Canada and by the Province of Ontario through 
the Ministry of Research, Innovation and Science. NA and MCJ are supported by the National Science and Engineering Research Council through a Discovery grant. This work was  performed in part at Aspen Center for Physics, which is supported by National Science Foundation grant PHY-1607611.

\end{acknowledgments}
\bibliography{Zero_Modes_ref}

\end{document}